\documentclass[%
reprint,
amsmath,amssymb,
aps,
]{revtex4-1}

\usepackage{graphicx}
\usepackage{dcolumn}
\usepackage{bm}
\usepackage{CJKutf8}

\begin{document}

\begin{CJK}{UTF8}{gbsn}

	\preprint{APS/123-QED}
	
\title{Impurity and vortex States in the bilayer
high-temperature superconductor La$_3$Ni$_2$O$_7$}

\author{Junkang Huang$^{1,2}$, Z. D. Wang$^3$, Tao Zhou$^{1,2,3}$}
\email{tzhou@scnu.edu.cn}

\affiliation{$^1$Guangdong Basic Research Center of Excellence for Structure and Fundamental Interactions of Matter, Guangdong Provincial Key Laboratory of Quantum Engineering and Quantum Materials, School of Physics, South China Normal University, Guangzhou 510006, China\\
	$^2$Guangdong-Hong Kong Joint Laboratory of Quantum Matter, Frontier Research Institute for Physics, South China Normal University, Guangzhou 510006, China\\
$^3$Guangdong-Hong Kong Joint Laboratory of Quantum Matter, Department of Physics, and HK Institute of Quantum Science \& Technology, The University of Hong Kong, Pokfulam Road, Hong Kong, China
}

\date{\today}

\begin{abstract}
We perform a theoretical examination of the local electronic structure in the recently discovered bilayer high-temperature superconductor La$_3$Ni$_2$O$_7$. Our method begins with a bilayer two-orbital tight-binding model, incorporating various pairing interaction channels. We determine superconducting order parameters by self-consistently solving the real-space Bogoliubov-de Gennes (BdG) equations, revealing a robust and stable extended s-wave pairing symmetry.
We investigate the single impurity effect using both self-consistent BdG equations and non-self-consistent T-matrix methods, uncovering low-energy in-gap states that can be explained with the T-matrix approach. Additionally, we analyze magnetic vortex states using a self-consistent BdG technique, which shows a peak-hump structure in the local density of states at the vortex center. Our results provide identifiable features that can be used to determine the pairing symmetry of the superconducting ${\mathrm{La}_3\mathrm{Ni}_2\mathrm{O}_7}$ material.

\end{abstract}

\maketitle

\section{\label{Intro}Introduction}

Superconductivity with a critical temperature (T$_c$) of approximately 80 K has been recently discovered in the bulk bilayer nickelate compound  $\mathrm{La}_3\mathrm{Ni}_2\mathrm{O}_7$ under pressures between 14.0-43.5 GPa~\cite{Sun2023}. Notably, the average valence of the nickel ions in this compound is $+2.5$, associated with the average $3d^{7.5}$ configuration—a significant departure from the traditionally observed $3d^9$ configuration in previous infinite-layer nickelate superconductors~\cite{Li2019}. This discovery positions the material as a novel platform for exploring high-T$_c$ superconductivity, consequently instigating a wave of both theoretical and experimental investigations~\cite{arXiv2305.15564,arXiv2306.03231,arXiv2306.03706,arXiv2306.06039,arXiv2306.07275,arXiv2307.10144,arXiv2307.15276,
arXiv2307.15706,arXiv2307.16873,zhang2023trends,tian2023correlation,qin2023hightc,arXiv2306.05121,arXiv2306.07837,arXiv2306.07931,arXiv2306.14841,
arXiv2307.05662,arXiv2307.06806,arXiv2307.14965,arXiv2307.16697,arXiv2308.01176,jiang2023high,arXiv2307.07154,arXiv2307.09865,arXiv2307.02950,arXiv2307.14819,s11433-022-1962-4}.
The low energy electronic structures of this compound, as determined through density-functional-theory calculation, are majorly contributed to by Ni-$3d_{z^2}$ and Ni-$3d_{x^2-y^2}$ orbitals. Consequently, the bilayer two-orbital model (incorporating four energy bands) is widely employed to obtain normal state energy bands~\cite{arXiv2305.15564,arXiv2306.03231,arXiv2306.03706,arXiv2306.06039,arXiv2306.07275,arXiv2307.10144,arXiv2307.15276,
arXiv2307.15706,arXiv2307.16873,arXiv2306.05121,zhang2023trends,tian2023correlation,qin2023hightc,arXiv2306.07837,arXiv2306.07931,arXiv2306.14841,
arXiv2307.05662,arXiv2307.06806,arXiv2307.14965,arXiv2307.16697,arXiv2308.01176,jiang2023high}.

It was indicated experimentally~\cite{arXiv2307.02950} and proposed theoretically ~\cite{arXiv2305.15564,arXiv2306.03231,arXiv2306.03706,arXiv2306.06039,arXiv2306.07275,arXiv2307.10144,arXiv2307.15276,
arXiv2307.15706,arXiv2307.16873,zhang2023trends,tian2023correlation,qin2023hightc,arXiv2306.05121,arXiv2306.07837,arXiv2306.07931,arXiv2306.14841,
arXiv2307.05662,arXiv2307.06806,arXiv2307.14965,arXiv2307.16697,arXiv2308.01176,jiang2023high} that strong electronic correlations exist in the $\mathrm{La}_3\mathrm{Ni}_2\mathrm{O}_7$ compound, implicating the likelihood of an unconventional pairing mechanism. Strategically identifying the pairing symmetry could be instrumental in studying the superconducting pairing mechanism. To date, the pairing symmetry has been investigated theoretically through diverse theoretical methods, leading to the prediction of multiple potential pairing symmetries, including several different $s_{\pm}$-wave pairing~\cite{arXiv2306.03706,arXiv2306.06039,arXiv2306.07275,arXiv2307.10144,arXiv2307.15276,arXiv2307.15706,arXiv2307.16873,zhang2023trends,tian2023correlation}, the $d$-wave pairing~\cite{arXiv2306.05121,jiang2023high}, and the coexistence/competition of the $s$-wave and $d$-wave pairing symemtries~\cite{arXiv2307.14965,arXiv2307.16697}. 
 This indicates that the chosen pairing symmetry could depend on the particular model and approximation considered. Hence, further experimental evidence is critical for definitive determination of the pairing symmetry. 
Theoretically, the Bogoliubov-de Gennes (BdG) equation based on real-space self-consistent calculations is a powerful method for obtaining the pairing symmetry of unconventional superconductors. In the past, this method was widely used to study the local electronic structure of high-temperature superconductors, and the pairing symmetry was correctly obtained through self-consistent calculations~\cite{PhysRevB.83.214502,PhysRevLett.87.147002,PhysRevB.80.014523,PhysRevB.80.134505}.

The local electronic structure is also a valuable method for studying the pairing symmetry of superconductors.
It can be explored using scanning tunneling microscopy experiments, which are effective in discerning the pairing symmetry by measuring the local electronic structure in proximity to an impurity or magnetic vortex~\cite{RevModPhys.79.353}. Theoretically, such local electronic structures can be examined through the calculation of the local density of states (LDOS). Past efforts to investigate the impurity and vortex-induced bound states have indeed provided valuable insights into the electronic structure and pairing symmetry of various unconventional superconductors~\cite{RevModPhys.79.353,RevModPhys.78.373,PhysRevLett.103.186402,PhysRevB.80.064513,Li2021,PhysRevB.83.214502,PhysRevLett.87.147002,PhysRevB.80.014523,PhysRevB.80.134505,PhysRevB.105.174518,PhysRevB.106.014501}.

In this paper, we provide a theoretical investigation into the LDOS in the presence of a single impurity or magnetic vortices in the superconducting $\mathrm{La}_3\mathrm{Ni}_2\mathrm{O}_7$ material, utilizing the BdG equations and a bilayer two-orbital model. Notably, our self-consistent calculations predict an extended $s$-wave pairing symmetry as the most favorable outcome, based on which, we calculate the LDOS and discuss the impurity and vortex induced low energy states. Furthermore, the impact of impurities is analyzed using the $T$-matrix method, offering a numerical analysis of the origin of the impurity-induced low energy states. We propose that different pairing symmetries can be distinguished through the impurity and vortex-induced bound states.

The remainder of this paper is structured as follows: In section II, we introduce the model and outline the pertinent formalism. Section III discusses the numerical results obtained from the self-consistent calculation, focusing on the impurity and vortex states in the La$_3$Ni$_2$O$_7$ compound. Section IV presents the numerical results from non-self-consistent calculations, offering a comparison to the self-consistent results from Section III. Finally, a concise summary is provided in Section V.

\section{\label{sec:Model}Model and Formalism}

We start with a bilayer two-orbital model to represent the normal state bands. Then we introduce the superconducting pairing term, derived from a phenomenological model that considers both nearest-neighbor intra-layer and onsite inter-layer attractive interactions, qualitatively consistent with the bilayer $t-J$-type model~\cite{arXiv2306.07275}. As a result, the comprehensive model can be represented as follows:
\begin{eqnarray}
H &=& \sum_{l,l^\prime}\sum_{{ij}\tau\tau^\prime\sigma} t^{l,l^\prime}_{{ij}\tau\tau^\prime} c^{l\dagger}_{{i}\tau\sigma} c^{l^\prime}_{{j}\tau^\prime\sigma}+ V_i\sum_{\tau\sigma} c^{1\dagger}_{{i_0}\tau\sigma} c^{1}_{{i_0}\tau\sigma} \nonumber\\
&&+\sum_{l,l^\prime}\sum_{{ ij}\tau} (\Delta^{l,l^\prime}_{{ij}\tau} c^{l\dagger}_{{i}\tau\uparrow} c^{l^\prime\dagger}_{{j}\tau\downarrow} + H.C.).
\end{eqnarray}
Here $l$ and $l^\prime$ are layer indices. $\tau$ and $\tau^\prime$ are orbital indices. ${i}$ and ${j}$ are site indices. $\sigma$ is the spin index. The normal state band parameters are taken from Ref.~\cite{arXiv2306.07275}. 

To study the point impurity effect, we consider a point impurity at the layer $1$ and the site ${i_0}$, with
$V_i$ being the impurity scattering strength. In the presence of the magnetic field, the hopping constants $t^{l,l^\prime}_{{\bf ij}\tau\tau^\prime}$ are replaced as $t^{l,l^\prime}_{{\bf ij}\tau\tau^\prime}\Rightarrow t^{l,l^\prime}_{{\bf ij}\tau\tau^\prime}\mathrm{exp}(i\phi_{\bf ij})$, where $\phi_{\bf ij}= \frac{\pi}{\Phi_0} \int_{\bf j}^{\bf i} A( {\bf r} ) \cdot d{\bf r}$. $\Phi_0$ is superconducting flux quantum and $A$ is vector potential in the Landau gauge with $A = \left( -By,0,0 \right)$.

The Hamiltonian can be diagonalized by solving the BdG equations,
\begin{eqnarray}
\sum_{l^\prime{j}\tau^\prime}\left(\begin{array}{cc}
{H^{l,l^\prime}_{{ij}\tau\tau^\prime,\sigma}}&{\Delta^{l,l^\prime}_{ij\tau}}\delta_{\tau,\tau^\prime}\\
{\Delta^{l,l^\prime *}_{ij\tau}}\delta_{\tau,\tau^\prime}&{-H^{l,l^\prime *}_{ij\tau\tau^\prime,\bar{\sigma}}}
\end{array}\right)
\left(\begin{array}{c}u^{l^\prime}_{nj\tau^\prime}\\v^{l^\prime}_{nj\tau^\prime} \end{array}\right) = E_n\left(\begin{array}{c}u^{l^\prime}_{nj\tau^\prime}\\v^{l^\prime}_{nj\tau^\prime} \end{array}\right),
\end{eqnarray}
with $H^{l,l^\prime}_{{ij}\tau\tau^\prime,\sigma}=t^{l,l^\prime}_{{ij}\tau\tau^\prime}+V_{i}\delta_{ij}\delta_{i,i_0}\delta_{l,1}$.

The pairing order parameter $\Delta^{l,l^\prime}_{ij\tau}$ are determined self-consistently, 
\begin{eqnarray}
\label{EQ:selfNNs}
{\Delta^{l,l^\prime}_{ij\tau}} = \frac{V^{l,l^\prime}_{ij\tau}}{4} \sum_n \left( u^l_{ni\tau} v^{l^\prime *}_{nj\tau} + u^{l^\prime}_{nj\tau} v^{l *}_{ni\tau} \right) \tanh \left( \frac{\beta E_n}{2} \right),
\end{eqnarray}
where $l=l^\prime$ and $l\neq l^\prime$ represent the intralayer and interlayer pairing, respectively.

The LDOS at the site $i$ of the layer $l$ can be calculated as,
\begin{eqnarray}
\rho^l_{i} = \sum_{n\tau} \left[ \left| u^l_{ni\tau} \right|^2 \delta \left( \omega - E_n \right) + \left| v^l_{ni\tau} \right|^2 \delta \left( \omega + E_n \right) \right],
\end{eqnarray}
where the $\delta$-function is expressed as $\delta(x)=\Gamma/[\pi (x^2+\Gamma^2)]$ with $\Gamma=0.005$.

The effect of a single impurity can be studied theoretically via an alternative method, namely, the $T$-matrix method~\cite{RevModPhys.78.373,PhysRevLett.103.186402,PhysRevB.80.064513,Li2021}. When overlooking the suppression of the superconducting pairing amplitude by the impurity, the superconducting Hamiltonian [Eq. (1)] can be transformed into the momentum space with $H=\sum_{\bf k} \Psi^\dagger({\bf k})\hat{M}({\bf k})\Psi({\bf k})$. Here, $\hat{M}({\bf k})$ represents an $8\times 8$ matrix while the column vector $\Psi({\bf k})$ is given by:

\begin{equation}
\Psi({\bf k})=\left( c^1_{{\bf k}1\uparrow},c^1_{{\bf k}2\uparrow},c^2_{{\bf k}1\uparrow},c^2_{{\bf k}2\uparrow},c^{1\dagger}_{-{\bf k}1\downarrow},c^{1\dagger}_{-{\bf k}2\downarrow},c^{2\dagger}_{-{\bf k}1\downarrow},c^{2\dagger} _{-{\bf k}2\downarrow}      \right)^T.
\end{equation}

We define the bare Green's function matrix for a clean system, with the elements being defined as, 
\begin{eqnarray}
\label{EQ:Gij0}
G_{ij0} ( {\bf k},\omega ) = \sum_n \frac{u_{in}({\bf k})u_{jn}^*({\bf k})}{\omega - E_n({\bf k}) + i\Gamma},
\end{eqnarray}
where ${u}_n({\bf k})$ and $E_n({\bf k})$ are eigenvectors and eigenvalues of the matrix $\hat{M}({\bf k})$, respectively. 

Considering a single impurity at the site $(0,0)$ of layer $1$, the $T$-matrix can be expressed as,
\begin{eqnarray}
\label{EQ:Tmatrix}
\hat{T} \left( \omega \right) = \frac{\hat{U}}{ \hat{I} - \hat{U}\hat{G}_0\left( \omega \right) },
\end{eqnarray}
where $\hat{I}$ is an $8\times 8$ identity matrix, and $U$ is a diagonal matrix with  four nonzero elements: $U_{11} = U_{22} = V_{i}$ and $U_{55} = U_{66} = -V_{i} $. 

The full Green's function can be expressed as
\begin{eqnarray}
\label{EQ:Grromega}
\hat{G}\left( {\bf r},{\bf r^\prime},\omega \right) =  \hat{G_0}\left( {\bf r},{\bf r^\prime},\omega \right) + \hat{G_0}\left( {\bf r},0,\omega \right) \hat{T}\left( \omega \right) \hat{G_0}\left( 0,{\bf r^\prime},\omega \right),
\end{eqnarray}
where $G_0\left( {\bf r},{\bf r^\prime},\omega \right)$ is the Fourier transform of $G_0\left( {\bf k},\omega \right)$ with
$G_0\left( {\bf r},{\bf r^\prime},\omega \right) = \frac{1}{N}\sum_k G_0\left( {\bf k},\omega \right) e^{i{\bf k}\cdot ( {\bf r} - {\bf r^\prime} )}$.

The LDOS at the layer $l$ and the site ${\bf r}$ can be calculated by the full Green's function
\begin{eqnarray}
\label{EQ:rhoTM}
\rho^l( {\bf r},\omega ) &=& -\frac{1}{\pi} \mathrm{Im} \sum_{p = 1}^{2} [G_{m+p,m+p}( {\bf r},{\bf r},\omega )+\nonumber\\&& G_{m+p+4,m+p+4}( {\bf r},{\bf r},-\omega )],
\end{eqnarray}
with $m=2(l-1)$.

In the present work, with the BdG method, the real space system size of each layer is chosen as $N=40\times 40$.
For the exploration of the magnetic vortex effect, we set $B = 2\Phi_0 / N$ and $V_i=0$. In studying the single impurity effect, we designate $B=0$ and $V_i=10$. The calculation of LDOS involves the consideration of a $30\times 30$ supercell.
We have performed numerical tests to confirm our general conclusions. The results remain qualitatively consistent across various system sizes (ranging from $20\times 20$ to $40\times 40$) and impurity strengths (with $3\leq V_i\leq 50$).

\section{Self-Consistent Calculation Results}

We begin with a discussion of the most viable pairing symmetry. This is determined based on a self-consistent calculation, taking into account an attractive intra-layer interaction among nearest neighbors, denoted as $V^{l,l}_{ij\tau}$, and an onsite attractive inter-layer interaction represented by $V^{1,2}_{ii\tau}$. These terms contribute to the generation of superconducting pairing. Numerical validation confirms the robustness and stability of the $s$-wave pairing symmetry solution with $\Delta^{l,l}_{i,i+\hat{x}\tau}=\Delta^{l,l}_{i,i+\hat{y}\tau}$, regardless of the parameters of $V^{1,2}_{ii\tau}$, $V^{l,l}_{ij\tau}$ and the initial input parameters. The parameters chosen for the ensuing results are $V^{1,2}_{ii1}=V^{1,2}_{ii2}=0.4$ and $V^{l,l}_{ij1}=V^{l,l}_{ij2}=0.8$, with $i$ and $j$ denoting nearest-neighbor sites. Note that the choice of pairing potentials ($V^{l,l^\prime}_{ij\tau}$) is for illustration purposes only. We have numerically verified that the presented results remain qualitatively similar when considering different pairing potentials.
The conclusions are robust and applicable to different situations, as they are not limited to a specific set of parameters.

Our numerical results also reveal that superconducting pairing within the $d_{z^2}$ orbital is dominant, underpinned by a significant interlayer pairing term. The interlayer order parameters exhibit an inverse sign compared to the intralayer parameters. Interlayer order parameters in the $d_{x^2-y2}$ channel are relatively minimal. These findings align qualitatively with preceding mean-field self-consistent calculations based on the $t-J$-type model~\cite{arXiv2306.07275}.

\subsection{Impurity states}

\begin{figure}
	\includegraphics[scale=0.22]{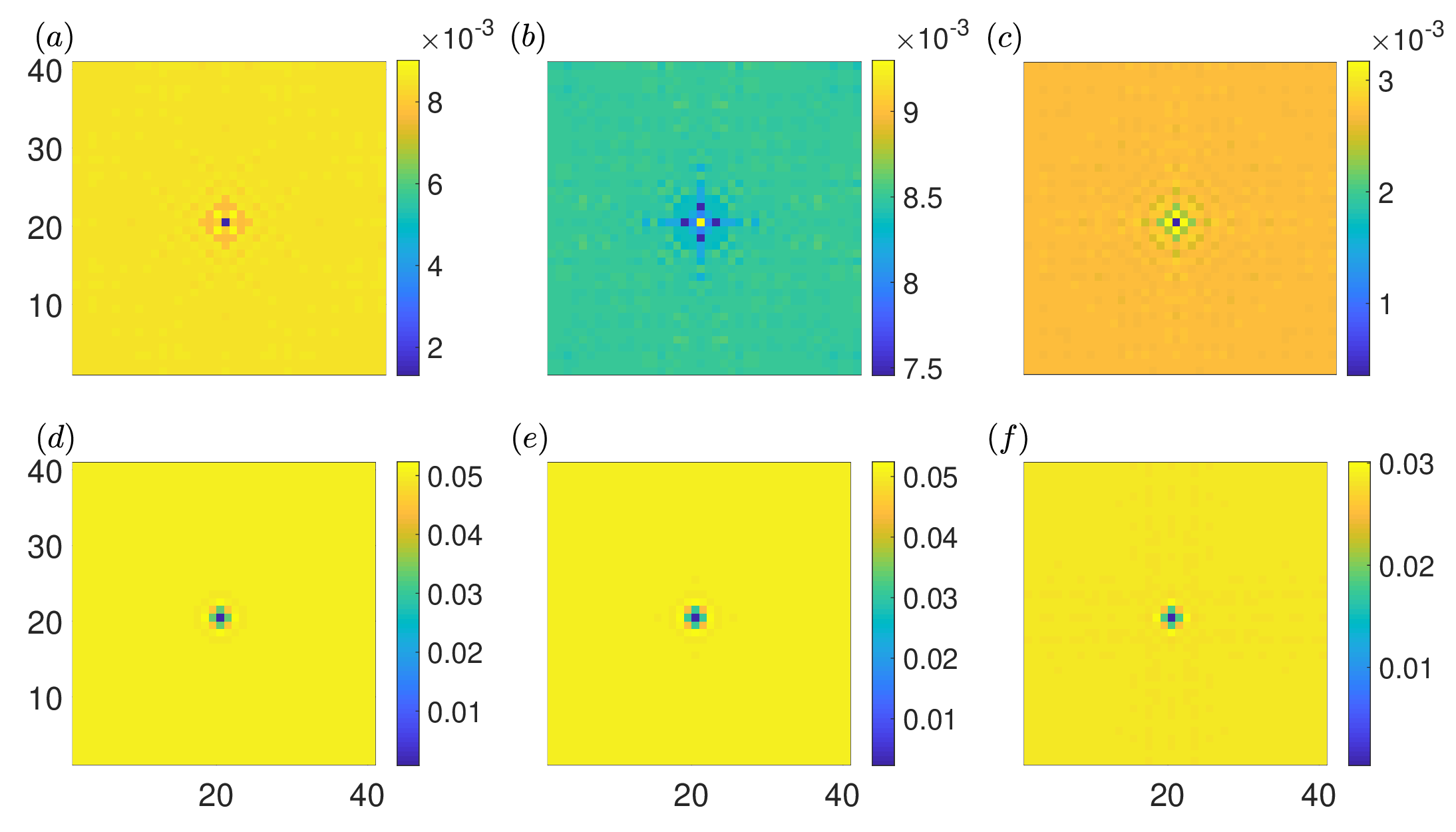}
	\caption{The site-dependent gap amplitudes [$\Delta^{l,l^\prime}_{\tau}(i)$] with a point impurity at the site (20, 20) of layer 1. The upper and lower panels depict the pairing amplitudes within the $d_{x^2-y^2}$ channel and $d_{z^2}$ channel, respectively. The panels on the left represent the intralayer pairing amplitudes for layer 1, while the middle panels depict the same for layer 2. On the right, the panels illustrate the interlayer pairing amplitudes.   }
	\label{fig.1}
\end{figure}

We now study the single impurity effect and discuss the possible impurity induced in-gap states with considering a point impurity at the site $(20,20)$ of layer 1. To perform this analysis, we solve the BdG equations [Eq.(2)], followed by the attainment of superconducting order parameters through a self-consistent method [Eq.(3)]. The site-dependent intralayer superconducting gap amplitudes, $\Delta^{l,l}_{\tau}(i)$, are specified as $\Delta^{l,l}_{\tau}(i)=\frac{1}{4}\mid \sum_j \Delta^{l,l}_{ij\tau}\mid$, with $j$ representing four nearest-neighbor sites of the site $i$. The spatial distribution of the order parameter amplitudes are plotted in Fig.~1. As is seen, the pairing amplitudes in the $d_{z^2}$ channel significantly exceed those within the $d_{x^2-y^2}$ channel. The amplitudes are suppressed almost to zero at the impurity site itself, but quickly escalate back to the bulk value as they move away from the impurity site.

Fig.~2 illustrates the effect of an impurity on LDOS spectra. For a strong point impurity, the electronic structure at the impurity site is significantly suppressed, resulting in zero LDOS. A closer examination of the LDOS spectra near the point impurity is presented in Figs. 2(a) and 2(b). These depict the spectra at site (20,20) of layer 2 and site (20,21) of layer 1, respectively. As a point of comparison, we also include the LDOS spectrum in the system bulk, where $V_i=0$, represented as dashed lines.
In the system bulk, the LDOS spectrum manifests a two-gap structure, consisting of two distinct gaps, with superconducting coherent peaks appearing at energies $\pm 0.05$ and $\pm 0.15$.  This two-gap feature is due to the multi-band effect.
Upon introducing an impurity, there is a noticeable upsurge in the low energy LDOS at the site above the impurity [site (20,20) on layer 2], indicating  impurity-induced in-gap states. A peak structure emerges at the positive energy, close to the first gap edge. At site (20,21) on layer 1, the effect of the impurity has been significantly mitigated with no evident peak structure resulting from the impurity. We have checked numerically that when moving further away from the impurity site, the LDOS spectrum aligns almost precisely with the system bulk spectrum.

\begin{figure}
	\includegraphics[scale=0.3]{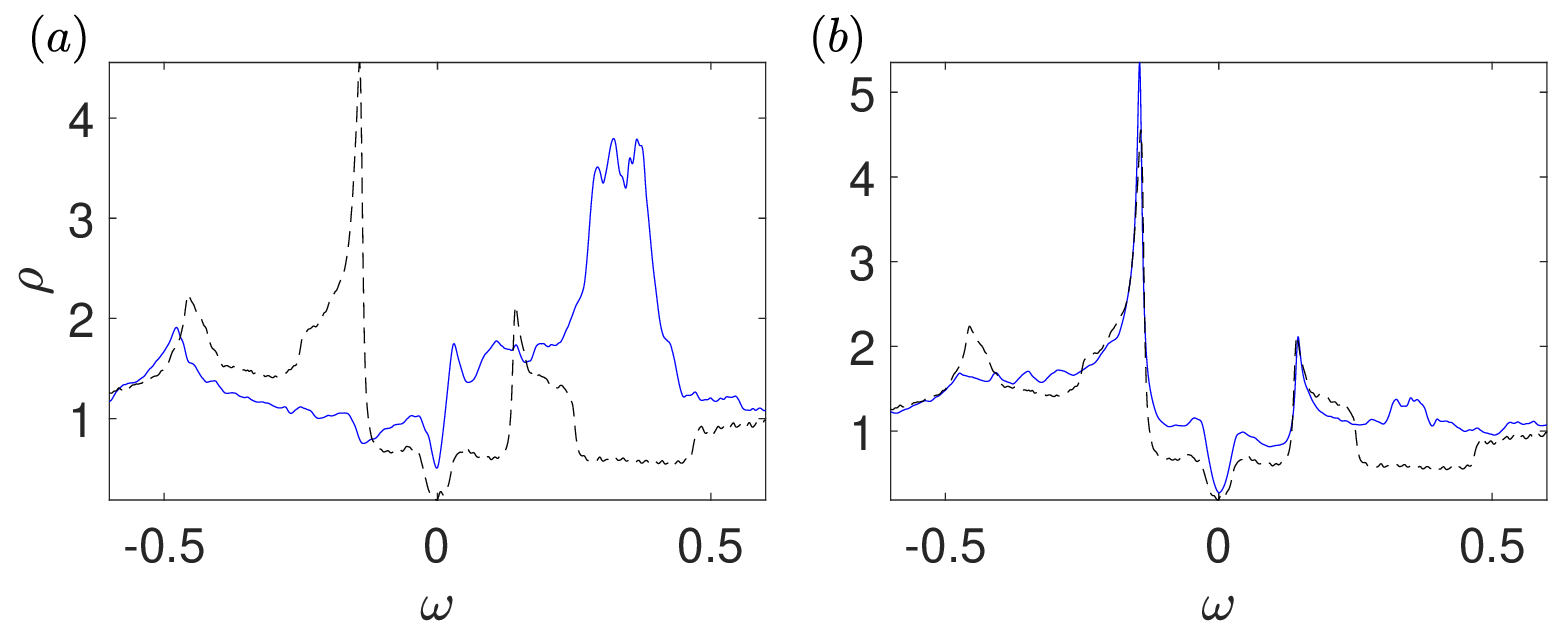}
	\caption{The LDOS as a function of the energy $\omega$ near the impurity site with the impurity strength $V_i=10$ (the solid lines) and $V_i=0$ (the dashed lines). (a) The LDOS at the upper site of the impurity site [site $(20,20)$ of layer 2]. (b) The LDOS at the right site of the impurity site [site $(21,20)$ of layer 1].} 
	\label{fig.2}
\end{figure}

By neglecting the impurity-induced suppression of the order parameter and assuming that the pairing order parameters are uniform, the point impurity effect can be studied with the $T$-matrix method~\cite{RevModPhys.78.373,PhysRevLett.103.186402,PhysRevB.80.064513,Li2021}. The $T$-matrix method is broadly accepted for its ability to capture the fundamental physics and accurately depict the single impurity effect on a qualitative level. We also employ the $T$-matrix method to examine the impurity effect and contrast the results with those obtained from the BdG technique.  Numerical results of the LDOS spectra in the proximity to an impurity site, generated through the $T$-matrix method, are presented in Fig. 3. These results are qualitatively consistent with those generated through the BdG technique. As previously discussed~\cite{PhysRevB.80.064513}, the primary distinction between the self-consistent BdG technique and the $T$-matrix method, in terms of the impurity effect, lies in the inclusion or exclusion of self-energy corrections induced by the impurity. Generally, these corrections are proportional to the impurity concentration and can be reasonably disregarded when studying the point impurity effect.

\begin{figure}
	\includegraphics[scale=0.3]{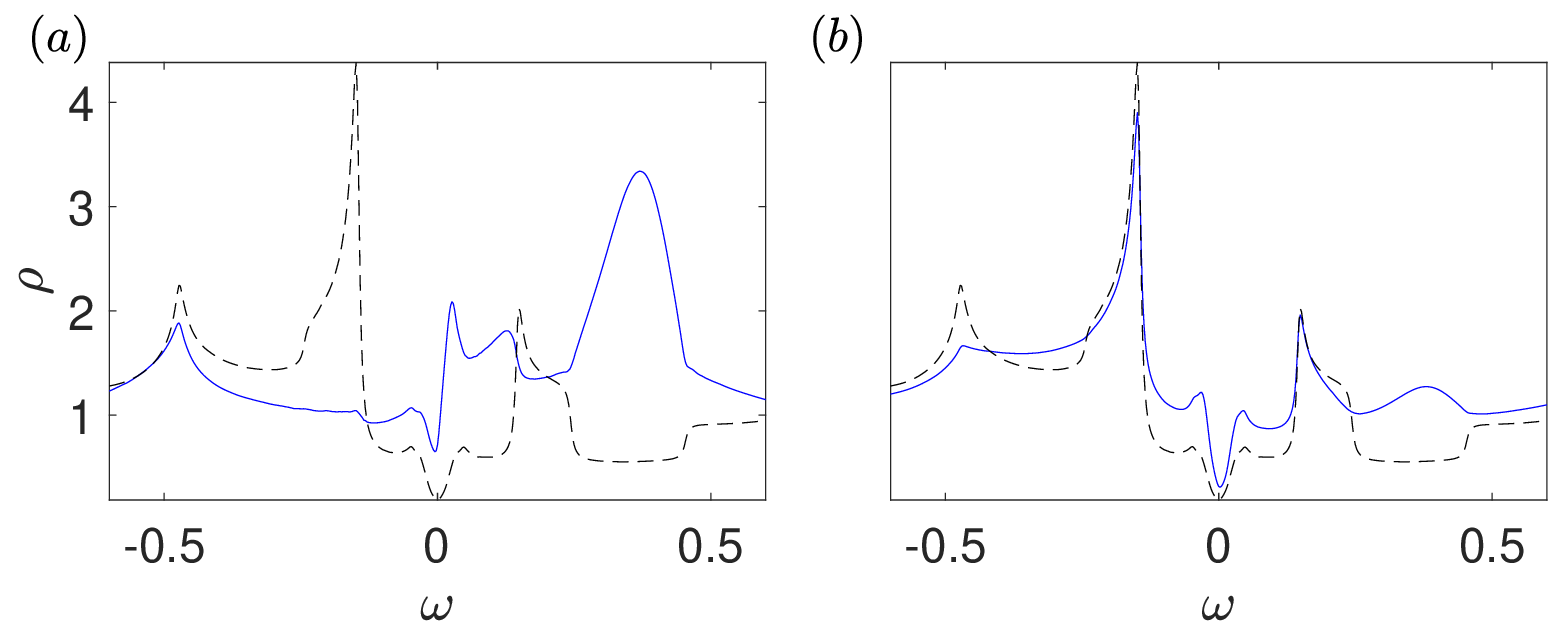}
	\caption{Similar to Fig.~2 but based on the T-matrix method.} 
	\label{fig.3}
\end{figure}

\begin{figure}
	\includegraphics[scale=0.5]{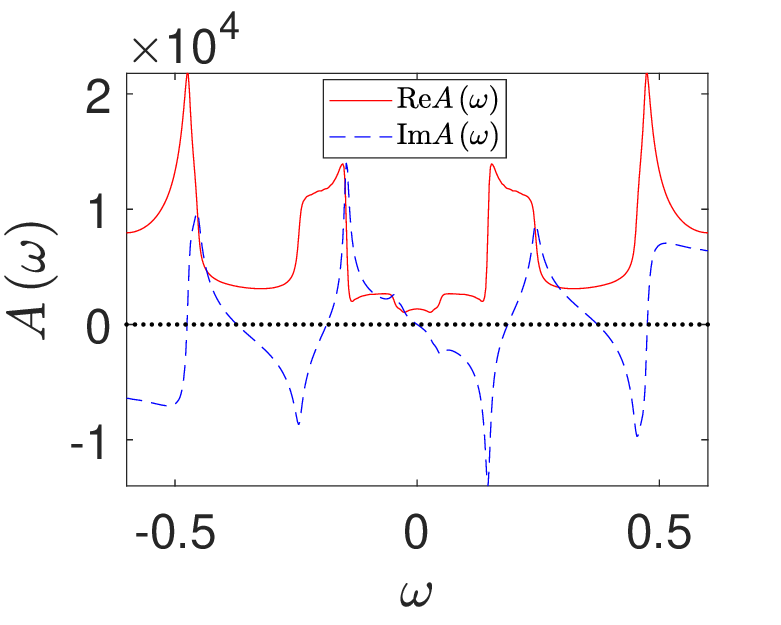}
	\caption{The real and imaginary parts of $A(\omega)$ as a function
		of the energy $\omega$ with $V_i=10$.} 
	\label{fig.4}
\end{figure}

The advantage of the $T$-matrix method lies in its ability to offer an in-depth understanding of in-gap states in LDOS spectra through analysis of the denominator of the $T$-matrix. This can be achieved by defining the complex function, $A(\omega)$, as $A(\omega)=\mathrm{Det} [\hat{I} - \hat{U}\hat{G}_0\left( \omega \right) ]$, where $\mathrm{Det}(\hat{M})$ denotes the determinant of the matrix $\hat{M}$. From $A(\omega)$, the bare Green's function $\hat{G_0}$ contributes to the imaginary part [Im$A(\omega)$].
In the superconducting state, due to the existence of the superconducting gap, Im$A(\omega)$ is typically small at low energies. If the real part of $A(\omega)$ is minimal or crosses the zero axis, it results in a large $T$-matrix. This subsequently causes low-energy in-gap states. The real and imaginary parts of $A(\omega)$, depicted as a function of energy $\omega$, are displayed in Fig. 4.
These provide a comprehensive understanding of the LDOS spectra. Specifically, when the real and imaginary parts of $A(\omega)$ are small at low energies, it leads to impurity-induced, low-energy states. Furthermore, the real part of $A(\omega)$ reaches its minimum at a specific low energy of approximately $\pm 0.05$. At this energy, the LDOS spectra near the impurity will significantly amplify, and a peak structure will emerge, as seen in Figs. 2 and 3.

\subsection{Magnetic vortex states}
\begin{figure}
	\includegraphics[scale=0.25]{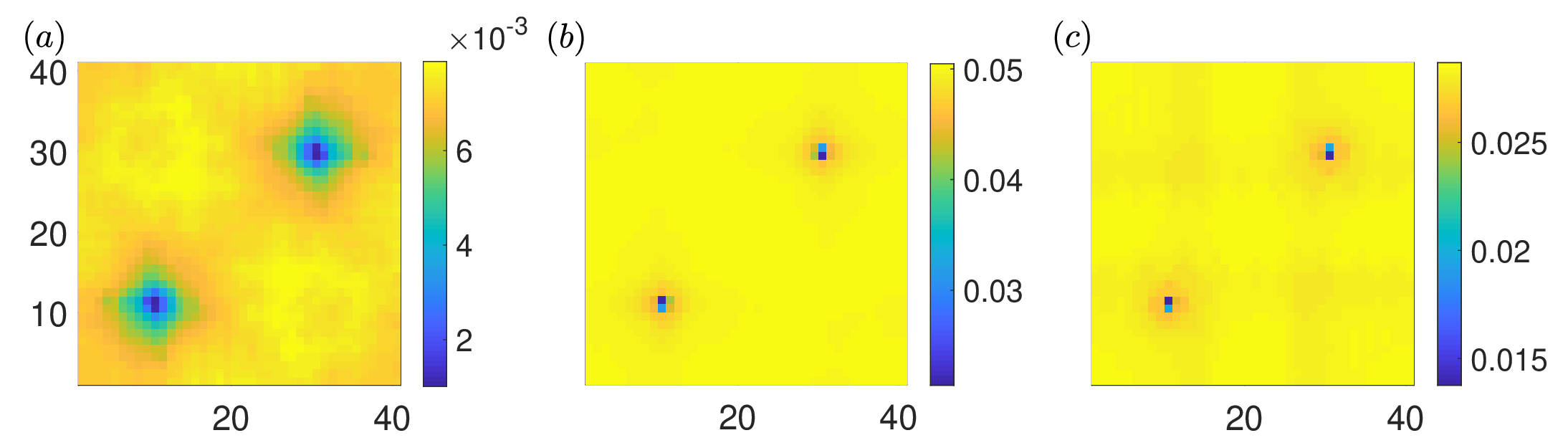}
	\caption{Intensity plots of the order parameter amplitudes in the presence of the vortices. (a) The intralayer pairing amplitude in the $d_{x^2-y^2}$ orbital channel. (b) The intralayer pairing amplitude in the $d_{z^2}$ orbital channel.
	(c) The interlayer pairing amplitude in the $d_{z^2}$ orbital channel. }
	\label{fig.5}
\end{figure}

We turn to study the magnetic vortex states. 
Fig.~5 shows the self-consistent results of the order parameter amplitudes in the presence of a magnetic field. It can be seen that the superconducting gap decreases as the magnetic field intensifies. In this situation, the interlayer pairing in the $d_{x^2-y^2}$ channel is insignificantly small and is therefore not showcased in this study.
In both the intralayer and interlayer pairing, two vortices are seen to emerge, with their centers located at sites $(10,11)$ and $(30,29)$. The order parameter ceases at the vortex center, gradually increases outward, and returns to the standard values at the vortex edges. The radius of the vortex represents the scale of the superconducting coherence length which depends heavily on the gap amplitudes. As the superconducting gap amplifies, the coherence length reciprocally diminishes.
As previously mentioned, in the system's bulk, the order parameter magnitudes in the $d_{z^2}$ orbital significantly overweight those in the $d_{x^2-y2}$ channel. Therefore, the superconducting coherence length should be smaller in the $d_{z^2}$ channel.
As a result, and as seen in Figs.~5(a)-(c), the vortex region in the $d_{x^2-y2}$ channel is more extensive than that of the $d_{z^2}$ channel.

Now let us discuss the electronic structure in the presence of the vortices. The LDOS spectra, as a function of energy $\omega$ from the bulk to the vortex center, are visualized in Fig.~6(a). As evident, a notable in-gap resonant peak at a specific negative energy (approximately -0.09) stands out. In addition to the strong in-gap peak at negative energy, the LDOS features a broad zero-energy bump structure, which subsequently forms a peak-hump configuration at the vortex center.
When moving away from the vortex center, the peak intensity dramatically decreases. Ultimately, the peak and hump structure capitulate, allowing the LDOS to revert to its standard bulk feature upon reaching a site that is three lattice constants distant from the vortex center.

\begin{figure}
	\includegraphics[scale=0.22]{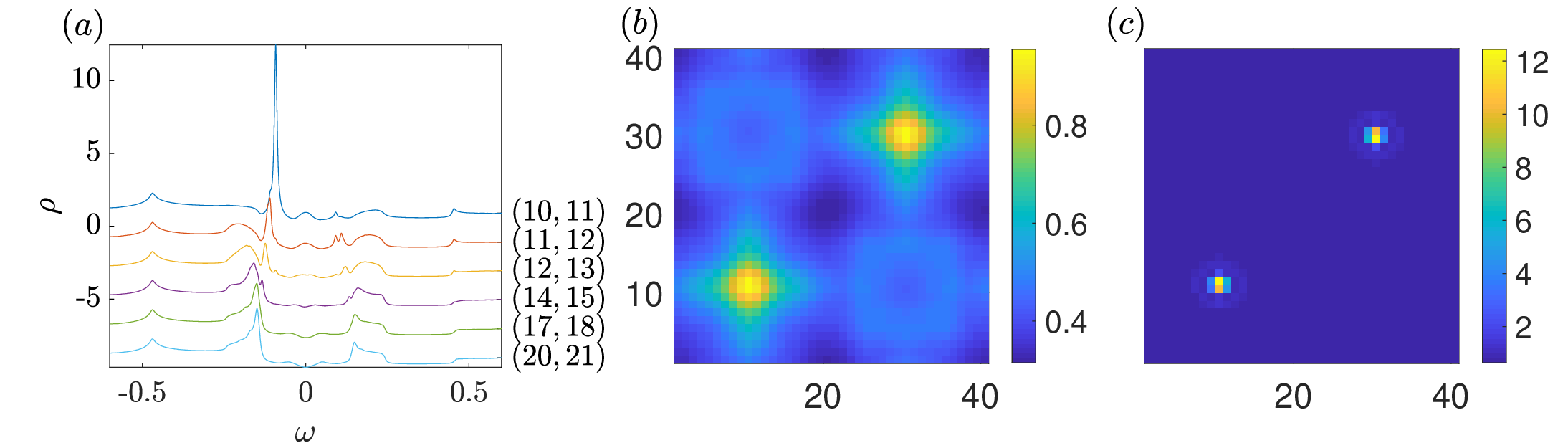}
	\caption{Numerical results of the LDOS in the presence of the magnetic field. (a) The LDOS as a function of the energy $\omega$. (b) The intensity plot of the LDOS in the real space at the constant energy with $\omega=0$. (c) The intensity plot of the LDOS in the real space at the constant energy with $\omega=-0.09$. }
	\label{fig.6}
\end{figure}

Further evidence of the vortex states can be demonstrated through the exploration of the spatial variation of the LDOS spectra at the hump energy $(\omega=0)$ and the peak energy $(\omega=-0.09)$, as depicted in Figs.~6(b) and 6(c). For both energies, the greatest LDOS is observed at the vortex center site. At zero energy, the LDOS spectrum presents a broad structure, which is mirrored with a broad spatial distribution of the zero-energy states.
Contrarily, at the resonant peak energy, the LDOS spectrum exhibits a sharp peak. Consequently, the spatial distribution of the vortex states at this energy is also sharp, predominantly emerging at the vortex center site.

\section{Non-Self-Consistent Calculation Results}

Previously, different pairing symmetries were proposed by other groups using different methods. In Ref.~\cite{arXiv2306.03706}, an on-site $s_{\pm}$ pairing was suggested based on the functional renormalization group calculation, where the order parameter phases are different for different orbitals. The $d_{x^2-y^2}$-wave pairing symmetry was also proposed based on the spin fluctuation scenario~\cite{arXiv2306.05121}. However, these two pairing symmetries cannot be obtained with the self-consistent mean-field method.

To study the local electronic structure for these two pairing symmetries, a non-self-consistent method is used. For investigating the magnetic vortex effect, the superconducting phases with two vortices are set by hand. These can be obtained by solving a conventional superconductor in the presence of a magnetic field self-consistently using the BdG technique. The $T$-matrix method is employed to study the single impurity effect.

By exploring the local electronic structure for the $s_{\pm}$ and $d_{x^2-y^2}$ pairing symmetries in a non-self-consistent way, we can gain insights into the characteristics of these symmetries and their influence on the superconducting properties. The magnetic vortex effect and single impurity effect can provide valuable information about the pairing symmetries that may not be directly obtainable through self-consistent mean-field methods.

\begin{figure}
	\includegraphics[scale=0.3]{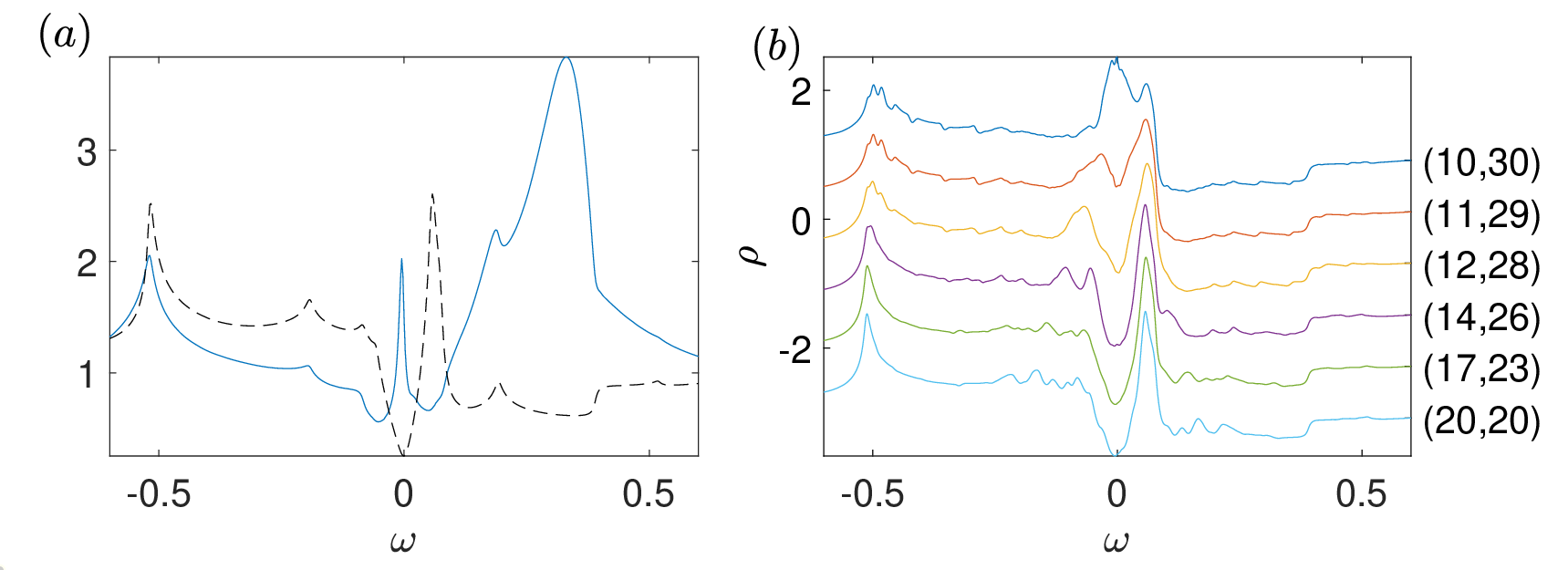}
	\caption{ The LDOS as a function of the energy $\omega$ for the $d_{x^2-y^2}$ pairing symmetry. (a) The solid line is the LDOS at the site $(0,0)$ of layer 2 by putting an impurity at the site $(0,0)$ of layer 1 with the impurity strength $V_i=10$. The dashed line is the bare LDOS without the impurity.  (b) The LDOS spectrum from the bulk to the vortex center (from the bottom to the up). }
\end{figure}

We first present the numerical results of the LDOS spectra for the $d_{x^2-y^2}$ pairing symmetry. The impurity effect is displayed in Fig.~7(a). In the system bulk, the bare LDOS spectrum also has a two-gap structure, similar to the case of the extended $s$-wave pairing symmetry shown in Fig.~2. In the presence of the impurity, a sharp zero energy resonance state emerges. This result is consistent with the single impurity effect for the $d$-wave pairing symmetry reported previously~\cite{RevModPhys.78.373}. 
The LDOS spectra in the presence of a magnetic field are plotted in Fig.~7(b). At the vortex center, which is the site (10,30), a broad peak appears near the zero energy. This finding provides further insights into the behavior of the $d_{x^2-y^2}$ pairing symmetry and its potential influence on the superconducting properties of the material.

We next present the numerical results of the LDOS spectra for the on-site $s_\pm$ pairing symmetry. Following the guidelines from Ref.~\cite{arXiv2306.03706}, both intralayer and interlayer pairing are considered. In this case, the signs of intralayer pairing in the $d_{z^2}$ and $d_{x^2-y^2}$ channels are opposite. Pairing amplitudes for distinct channels are sourced from Ref.~\cite{arXiv2306.03706}.

Fig.~8(a) shows the LDOS spectrum near an impurity. A slight increase in the LDOS at lower energies indicates the induction of low-energy states by the impurity, a characteristic significantly different from the extended $s$-wave pairing. Notably, here the impurity does not induce a peak. Near the gap edge, the intensity of the LDOS spectrum decreases.

\begin{figure}
	\includegraphics[scale=0.3]{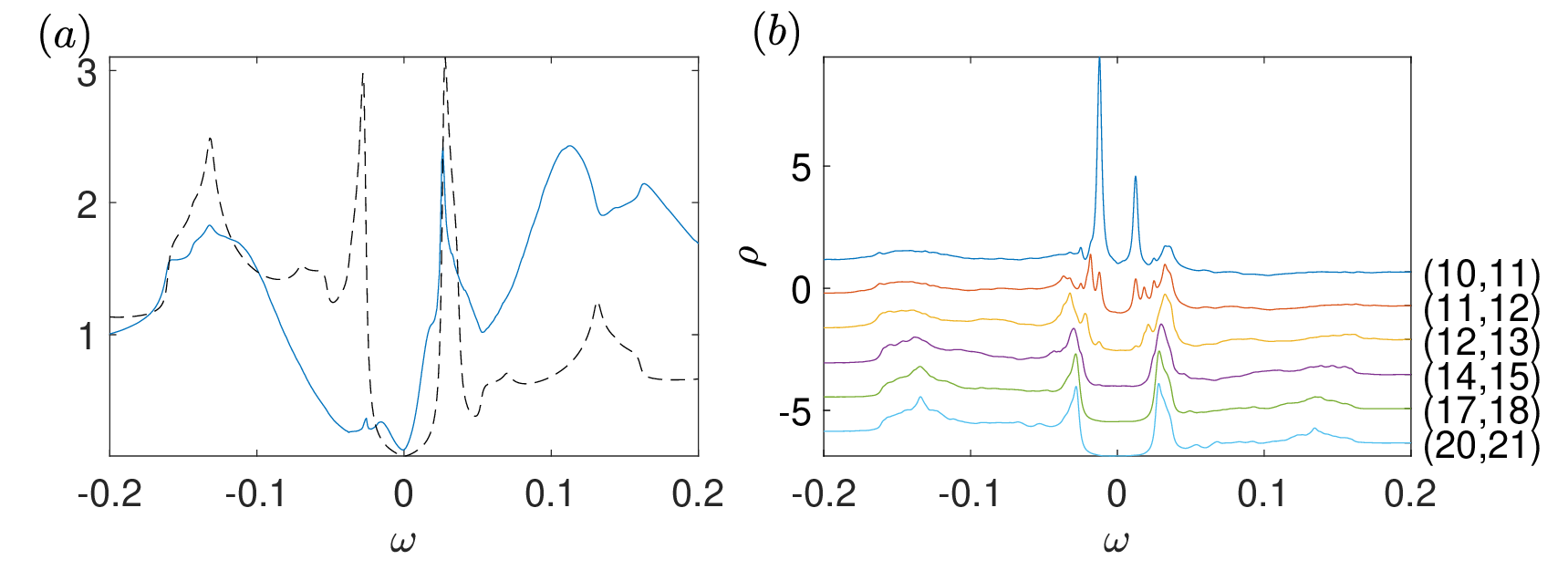}
	\caption{ Similar to Fig.~7 but for the onsite $s_{\pm}$ pairing symmetry. }
\end{figure}

In the presence of a magnetic field, Fig.~8(b) depicts LDOS spectra ranging from the bulk to the vortex center. A pair of in-gap peaks are induced at the vortex center [site (10,11)], with their positions symmetrically surrounding the Fermi energy. As the distance from the vortex center increases, the peak intensities gradually reduce, and the spectrum eventually evolves to resemble the bulk LDOS.

The characteristics of the local electronic structure near an impurity or crossing a vortex for both $d$-wave pairing symmetry and on-site $s_\pm$ pairing symmetry are significantly different from those of the previously discussed extended $s$-wave pairing symmetry. Therefore, it can be inferred that the effects of a single impurity and the magnetic vortex states can indeed provide valuable insights for identifying the pairing symmetry of the superconducting $\mathrm{La}_3\mathrm{Ni}_2\mathrm{O}_7$ material.

\section{summary}

In summary, a two-orbital ($d_{x^2-y^2}$ and $d_{z^2}$ orbitals) model featuring various channels of pairing potentials was employed to evaluate the physical attributes of the recently discovered high-T$_c$ superconducting compound $\mathrm{La}_3\mathrm{Ni}_2\mathrm{O}_7$. Based on real space self-consistent BdG equations, we suggest that the extended $s$-wave pairing should be the most advantageous pairing symmetry.  Furthermore, the calculations reveal that the superconducting gap magnitude within the $d_{z^2}$ orbital is significantly larger than that within the $d_{x^2-y^2}$ orbital. This indicates that the $d_{z^2}$ orbital plays a crucial role in determining the pairing symmetry and has a notable impact on the stability of the superconducting state in $\mathrm{La}_3\mathrm{Ni}_2\mathrm{O}_7$.

Both the BdG technique and the $T$-matrix method were utilized to evaluate the impurity effect. The computed LDOS spectra near the impurity site exhibited qualitative similarities. Additionally, the impurity propagated the in-gap low-energy states and led to the emergence of an in-gap peak at finite energy.
The magnetic vortex effect was also explored via the self-consistent BdG equations. The LDOS spectrum at the vortex center spotlights a peak-hump structure, with a pronounced peak at the negative energy dimensions and an amplified hump at zero energy. 

We also extend our study by taking into account two different pairing symmetries utilizing a non-selfconsistent method to investigate the point impurity effect and vortex states. The results obtained show significant qualitative differences compared to those based on the self-consistent calculation. Consequently, our numerical results suggest that the impurity effect and vortex states can be effectively utilized as a tool for investigating the pairing symmetry of the superconducting $\mathrm{La}_3\mathrm{Ni}_2\mathrm{O}_7$ material.

\begin{acknowledgments}
	This work was supported by the NSFC (Grant No.12074130), the Natural Science Foundation of Guangdong Province (Grant No. 2021A1515012340), the Key-Area Research and
Development Program of Guangdong Province (Grant No.2019B030330001), and the CRF of Hong Kong (C6009-20G).
\end{acknowledgments}


\providecommand{\noopsort}[1]{}\providecommand{\singleletter}[1]{#1}%

\end{CJK}

\end{document}